\documentclass[12pt,a4paper]{article}

\usepackage{geometry}
\usepackage{setspace}
\usepackage{overpic}
\usepackage{amsmath,amssymb}
\usepackage{natbib}

\newcommand{\Fig}[1]{Fig~\ref{fig:#1}}
\newcommand{\Eq}[1]{Eq~(\ref{eq:#1})}
\newcommand{\Dp}[2]{\frac{\partial #1}{\partial #2}}
\newcommand{\Sec}[1]{Sec~\ref{sec:#1}}
\newcommand{\dr}{\mathrm{d}}
\newcommand{\Dr}{\mathrm{D}}
\newcommand{\er}{\mathrm{e}}
\newcommand{\vs}{v_\mathrm{s}}
\newcommand{\kp}{k_\mathrm{p}}
\newcommand{\omp}{\omega_\mathrm{p}}
\newcommand{\alp}{\alpha}
\newcommand{\erfc}{\mathrm{erfc}}
\newcommand{\Ei}{\mathrm{Ei}}

\title{A Stokes drift approximation based on the Phillips
spectrum\footnote{Final version published in \textit{Ocean Modell}, 2016,
doi:10.1016/j.ocemod.2016.01.005}}

\author{{\O}yvind Breivik\thanks{Corresponding author. E-mail:
\texttt{oyvind.breivik@met.no}; ORCID Author ID: \texttt{0000-0002-2900-8458}.
Norwegian Meteorological Institute, All\'{e}g 70, NO-5007 Bergen, Norway}, 
   \and Jean-Raymond Bidlot\thanks{European Centre for Medium-Range Weather
         Forecasts (ECMWF).}
   \and Peter A.E.M. Janssen\footnotemark[3]}

\begin{document}
\maketitle

\begin{abstract}
A new approximation to the Stokes drift velocity profile based on the exact
solution for the Phillips spectrum is explored. The profile is compared with
the monochromatic profile and the recently proposed exponential integral
profile.  ERA-Interim spectra and spectra from a wave buoy in the central
North Sea are used to investigate the behaviour of the profile. It is found
that the new profile has a much stronger gradient near the surface and lower
normalized deviation from the profile computed from the spectra. Based on
estimates from two open-ocean locations, an average value has been estimated
for a key parameter of the profile. Given this parameter, the profile can
be computed from the same two parameters as the monochromatic profile,
namely the transport and the surface Stokes drift velocity.

Keywords: Stokes drift; Wave modelling; Stokes-Coriolis force; Langmuir
turbulence parameterization; Trajectory modelling.
\end{abstract}
\clearpage

\section{Introduction}
The Stokes drift  \citep{sto47} is defined as the difference between the
Eulerian velocity in a point and the average Lagrangian motion of a particle
subjected to the orbital motion $\mathbf{u}_\mathrm{w}$ of a wave field,
\begin{equation}
   \mathbf{v}_\mathrm{s} = \left\langle \int^t  \mathbf{u}_\mathrm{w} \, \dr t 
       \cdot \nabla \mathbf{u}_\mathrm{w} \right\rangle.
   \label{eq:stokes}
\end{equation}
Here the averaging is over a period appropriate for the frequency of surface
waves \citep{leibovich83}. The Stokes drift velocity profile is required
for a number of important applications in ocean modelling, such as the
computation of trajectories of drifting objects, oil and other substances (see
\citealt{mcwilliams00}, \citealt{bre12b}, \citealt{rohrs12}, \citealt{rohrs15}
and references in \citealt{bre13}). Its magnitude and direction is required
for the computation of the Stokes-Coriolis force which enters the momentum
equation in Eulerian ocean models (\citealt{hasselmann70}, \citealt{weber83},
\citealt{jenkins87b}, \citealt{mcwilliams99}, \citealt{janssen04},
\citealt{polton05}, \citealt{janssen12}, and \citealt{bre15b}),
\begin{equation}
   \frac{\Dr\mathbf{u}}{\Dr t} = -\frac{1}{\rho_\mathrm{w}} \nabla p 
  + (\mathbf{u} + \mathbf{v}_\mathrm{s}) \times f\hat{\mathbf{z}}
  + \frac{1}{\rho_\mathrm{w}} \frac{\partial \boldsymbol{\tau}}{\partial z}.
  \label{eq:stcor}
\end{equation}
Here $\mathbf{u}$ is the Eulerian current vector, $f$ the Coriolis frequency,
$\rho_\mathrm{w}$ the density of sea water, $\mathbf{v}_\mathrm{s}$ the
Stokes drift velocity vector, $\hat{\mathbf{z}}$ the upward unit vector,
$p$ the pressure and $\boldsymbol{\tau}$ the stress.

Langmuir circulation, first investigated by \citet{langmuir38}, manifests
itself as convergence streaks on the sea surface roughly aligned with the
wind direction. In a series of papers (\citealt{craik76}, \citealt{craik77},
\citealt{leibovich77}, \citealt{leibovich80}) a possible instability
mechanism arising from a vortex force $\mathbf{v}_\mathrm{s} \times
\boldsymbol{\omega}$ between the Stokes drift and the vorticity of the
Eulerian current was proposed to explain the phenomenon (named the second
Craik-Leibovich mechanism, CL2, by \citealt{faller78}). It is now commonly
accepted that CL2 is the main cause of Langmuir circulation in the open
ocean \citep{thorpe04}.  Langmuir turbulence is believed to be important
for the formation and depth of the ocean surface boundary layer (OSBL)
(\citealt{li97} and \citealt{flor10}), and a realistic representation
of the phenomenon in ocean models is important (see \citealt{axell02},
\citealt{rascle06}).  A common parameterisation  of the Langmuir turbulence
production term in the turbulent kinetic energy equation relates it to the
shear of the Stokes drift profile (\citealt{sky95}, \citealt{mcwilliams97},
\citealt{thorpe04}, \citealt{kantha04}, \citealt{ardhuin06}, \citealt{grant09}
and \citealt{belcher12}),
\begin{equation}
    \frac{\Dr e}{\Dr t} = 
   \nu_\mathrm{m} S^2 
    -\nu_\mathrm{h} N^2 
   +\nu_\mathrm{m} \mathbf{S} \cdot \Dp{\mathbf{v}_\mathrm{s}}{z} 
   -\Dp{}{z}(\overline{w'e}) 
   -\frac{1}{\rho_\mathrm{w}}\Dp{}{z}(\overline{w'p'})
   -\epsilon.
     \label{eq:tke}
\end{equation}
Here, $e$ represents the turbulent kinetic energy per unit mass; $w'e'$
and $w'p'$ are the turbulent transport and pressure correlation terms
(\citealt{stu88}, \citealt{kantha00}). The shear production and the
buoyancy production terms are well known quantities where $\mathbf{S} \cdot
\mathbf{S} = S^2 = (\partial \overline{\mathbf{u}}/\partial z)^2$, and $N^2 =
-(g/\rho_\mathrm{w}) \dr\rho_\mathrm{w}/\dr z $. Further, $\nu_\mathrm{h,m}$
are turbulent diffusion coefficients and $\epsilon$ represents the dissipation
of turbulent kinetic energy. It is the term  $\nu_ \mathrm{m} \mathbf{S}
\cdot \partial{\mathbf{v}_\mathrm{s}}/\partial{z}$, representing the Langmuir
turbulence production, that is of interest in this study. It is important to
note that it involves the shear of the Stokes drift. This quantity drops off
rapidly with depth, and clearly any parameterisation of the Langmuir production
term will depend heavily on the form of the Stokes drift velocity profile.

Climatologies of the surface Stokes drift have
been presented, either based on wave model integrations
\citep{rascle08,webb11,tamura12,rascle13,carrasco14,webb15} or on assumptions
of fully developed sea \citep{mcwilliams99}. However, the Stokes profile is
not so readily available as it is  expensive and impractical to integrate
the two-dimensional wave spectrum at every desired vertical level. It is also
numerically challenging to pass the full two-dimensional spectrum for every
grid point of interest from a wave model to an ocean model.  As was discussed
by \citet{bre14}, hereafter BJB, it has been common to replace the full Stokes
drift velocity profile by a monochromatic profile [see e.g.  \citet{sky95},
\citet{mcwilliams00}, \citet{car05}, \citet{polton05}, \citet{saetra07},
and \citet{tamura12}]. But this will lead to an underestimation of
the near-surface shear and an overestimation of the deep Stokes drift
\citep{ardhuin09,webb15}. This was partly alleviated by the exponential
integral profile proposed by BJB, but it too exhibited too weak shear near
the surface.

Here we explore a new approximation to the full Stokes drift velocity profile
based on the assumption that the Phillips spectrum \citep{phillips58} provides
a reasonable estimate of the intermediate to high-frequency part of the real
spectrum. The paper is organized as follows. First we present the proposed
profile in \Sec{theory}. We then investigate its behaviour for a selection
of parametric spectra in \Sec{paramspec} before looking at its performance
on two-dimensional wave model spectra in \Sec{era} for two locations
with distinct wave climate, namely the North Atlantic and near Hawaii.
The latter location is swell-dominated whereas the former exhibits a mix
of swell and wind sea \citep{rei11,semedo15} typical of the extra-tropics.
Finally, in \Sec{conc} we discuss the results and we present our conclusions
along with some considerations of the usefulness of the proposed profile
for ocean modelling and trajectory estimation.

\section{Approximate Stokes drift velocity profiles}
\label{sec:theory}
For a directional wave spectrum $E(\omega,\theta)$ the Stokes
drift velocity in deep water is given by
\begin{equation}
   \mathbf{v}_\mathrm{s}(z) = \frac{2}{g} \int_0^{2\pi}\!\int_0^{\infty}
   \omega^3 \hat{\mathbf{k}} \er^{2kz}
    E(\omega,\theta) \, \dr \omega \dr\theta,
   \label{eq:uvfom}
\end{equation}
where $\theta$ is the direction in which the wave component is travelling,
$\omega$ is the circular frequency and $\hat{\mathbf{k}}$ is the unit
vector in the direction of wave propagation.  This can be derived from the
expression for a wavenumber spectrum in arbitrary depth first presented by
\citet{kenyon69} by using the deep-water dispersion relation $\omega^2 =
gk$. For simplicity we will now investigate the Stokes drift profile under
the one-dimensional frequency spectrum
\begin{displaymath}
   F(\omega) \equiv \int_0^{2\pi} E(\omega,\theta) \dr \theta,
\end{displaymath}
for which the Stokes drift speed is written
\begin{equation}
   \vs(z) = \frac{2}{g}\int_0^{\infty} \omega^3F(\omega)\er^{2kz} \, \dr \omega.
   \label{eq:vs}
\end{equation}
From \Eq{vs} it is clear that at the surface the Stokes drift is proportional
to the third spectral moment [where the $n$-th spectral moment of the circular
frequency is defined as $m_n = \int_0^\infty \omega^n F(\omega) \, \dr\omega$],
\begin{equation}
   v_0 = 2m_3/g.
   \label{eq:v0}
\end{equation}

A new approximation to the Stokes drift profile was proposed by BJB,
and named the exponential integral profile,
\begin{equation}
   v_\mathrm{e} = v_0 
    \frac{\er^{2k_\mathrm{\er}z}}{1-Ck_\mathrm{\er}z},
    \label{eq:ve}
\end{equation}
where the constant $C=8$ was found to give the closest match. Here, the
inverse depth scale $k_\mathrm{\er}$ serves the same purpose as the average
wavenumber $k_\mathrm{m}$ used for a monochromatic profile,
\begin{equation}
   v_\mathrm{m} = v_0 \er^{2k_\mathrm{m}z}.
   \label{eq:vmono}
\end{equation}
The profile (\ref{eq:ve}) was found to be a much better approximation
than the monochromatic profile (\ref{eq:vmono}) with a 60\% reduction in
root-mean-square error reported by BJB, and has been implemented in the
Integrated Forecast System (IFS) of the European Centre for Medium-Range
Weather Forecasts (ECMWF); see \citet{janssen13} and \citet{bre15b}.

Here we propose a profile based on the assumption that the Phillips spectrum
\citep{phillips58}
\begin{equation}
   F_\mathrm{Phil} = \left\{ \begin{array}{lr}
             \alp g^2 \omega^{-5}, & \omega > \omp \\
             0,                                 & \omega \leq \omp 
                          \end{array} \right.,
   \label{eq:phil}
\end{equation}
yields a reasonable estimate of the part of the spectrum which contributes most
to the Stokes drift velocity near the surface, i.e., the high-frequency waves.
Here $\omega_\mathrm{p}$ is the peak frequency. We assume Phillips' parameter
$\alp = 0.0083$. The Stokes drift velocity profile under (\ref{eq:phil}) is
\begin{equation}
   v_\mathrm{Phil}(z) = 2\alp g \int_{\omp}^\infty 
       \omega^{-2} \er^{2\omega^2z/g} \,\dr\omega.
   \label{eq:vphil}
\end{equation}
An analytical solution exists for (\ref{eq:vphil}), see BJB, Eq~(11), which
after using the deep-water dispersion relation can be written as
\begin{equation}
   v_\mathrm{Phil}(z) = \frac{2\alp g}{\omp}
   \left[\exp{(2\kp z)} -
    \sqrt{-2\pi \kp z}\,\erfc \left(\sqrt{-2\kp z}\right)
     \right].
     \label{eq:vphilsolved}
\end{equation}
Here $\erfc$ is the complementary error function and $\kp = \omp^2/g$ is the
peak wavenumber. From (\ref{eq:vphilsolved}) we see that for the Phillips
spectrum (\ref{eq:vphil}) the surface Stokes drift velocity is
\begin{equation}
   v_0 \equiv v_\mathrm{Phil}(z=0) = \frac{2\alp g}{\omp}.
   \label{eq:v0p}
\end{equation}
For large depths, i.e. as $z \to -\infty$, \Eq{vphilsolved} approaches the
asymptotic limit [see BJB, Eqs~(14)-15)]
\begin{equation}
   \lim_{z \to -\infty} v_\mathrm{Phil} = 
   -\frac{v_0}{4\kp z} \er^{2\kp z}.
     \label{eq:limvz}
\end{equation}
This means the exponential integral profile (\ref{eq:ve}) proposed by BJB
has too strong deep flow when fitted to the Phillips spectrum. This could be
alleviated by setting the coefficient $C=4$ in \Eq{ve}, but at the expense of
increasing the overall root-mean-square (rms) deviation over the water column.
Further, although the profile (\ref{eq:ve}) is well suited to modelling the
shear at intermediate water depths, its shear near the surface is too weak.
Under the Phillips spectrum (\ref{eq:vphil}) the shear is
\begin{equation}
   \Dp{v_\mathrm{Phil}}{z} = 
     4 \alp \int_{\omp}^\infty
       \er^{2\omega^2z/g} \,\dr\omega,
   \label{eq:philshear}
\end{equation}
for which an analytical expression exists [see \citet{gradshteyn07},
Eq~(3.321.2)],
\begin{equation}
   \Dp{v_\mathrm{Phil}}{z} = 
     \alp \sqrt{-\frac{2\pi g}{z}} \,
     \erfc \left(\sqrt{-2\kp z}\right).
   \label{eq:philshear2}
\end{equation}
Near the surface the shear tends to infinity. This strong shear is not
captured by either the exponential integral profile (\ref{eq:ve}) or the
monochromatic profile (\ref{eq:vmono}).

Let us now assume that the Phillips spectrum profile (\ref{eq:vphilsolved})
is also a reasonable approximation for Stokes drift velocity profiles
under a general spectrum, and that the low-frequency part below the peak
contributes little to the overall Stokes drift profile so that it can be
ignored.  The general profile (\ref{eq:vs}) can be integrated by parts,
and for convenience we introduce the quantity
\begin{equation}
   G(\omega) = \int \omega^3 F(\omega) \, \dr \omega + C_1,
   \label{eq:anti}
\end{equation}
where $C_1$ is a constant of integration.
The integral (\ref{eq:vs}) can now be written
\begin{equation}
   \vs(z) = \frac{2}{g}\left(-G(\omega_\mathrm{p}) \er^{2\omp^2 z/g} -
                              \frac{4z}{g}\int_{\omega_\mathrm{p}}^{\infty}
                                \omega G(\omega)\er^{2\omega^2z/g} \,
                                \dr \omega\right).
   \label{eq:partial}
\end{equation}
We note that for the Phillips spectrum (\ref{eq:phil}), the quantity $\omega
G(\omega)$ becomes
\begin{equation}
   \omega G_\mathrm{Phil}(\omega) = \omega \left[\int_{\omp}^\omega s^3 F_\mathrm{Phil}(s) \, \dr s + C_1\right] =
   -\alp g^2 + \alp g^2\frac{\omega}{\omp} + C_1\omega,
   \label{eq:beta1}
\end{equation}
which is a constant, $-\alp g^2$, if we set $C_1 = -\alp g^2/\omp$. In this
case the solution to \Eq{partial} is \Eq{vphilsolved} as would be expected.

Assume now that in the range $\omp < \omega < \infty$ the quantity $\omega
G(\omega)$ is quite flat also for an arbitrary spectrum, and that it drops
to zero below $\omp$. Introduce
\begin{displaymath}
   \beta = -\frac{\langle \omega G(\omega) \rangle}{m_3 \omp},
\end{displaymath}
where the averaging operator is defined over a range of frequencies,
$\Delta \omega$, from the peak frequency to a cutoff frequency,
$\omega_\mathrm{c}$, such that $\langle X \rangle \equiv {\Delta \omega}^{-1}
\int_{\omp}^{\omega_\mathrm{c}} X \, \dr \omega $. Since we have  assumed
$\beta$ to be constant the $\omega G(\omega)$ in the second term of
\Eq{partial} can be factored out and we can approximate  \Eq{partial}
by \Eq{vphilsolved},
\begin{equation}
   \vs(z) \approx v_0
   \left[\er^{2\kp z} - \beta \sqrt{-2\kp \pi z} \, \erfc 
        \left(\sqrt{-2\kp z}\right) \right].
     \label{eq:vz}
\end{equation}
We note that if $F\,$ is the Phillips spectrum (\ref{eq:phil}) then
\begin{equation}
   \langle \omega G_\mathrm{Phil}(\omega) \rangle = -\langle \omega^5
   F_\mathrm{Phil}(\omega) \rangle = -\alpha g^2.
     \label{eq:omg}
\end{equation}
Assuming this to be a reasonable approximation for general spectra we find that we can approximate $\beta$ as follows,
\begin{equation}
   \hat{\beta} = \frac{2\langle \omega^5 F(\omega) \rangle}{g v_0 \omp}.
   \label{eq:hatbeta}
\end{equation}
Here we have substituted $m_3 = 2v_0/g$.  The Stokes transport $V =
\int_{-\infty}^0 v \, \dr z$ under \Eq{vz} can be found [see \ref{sec:philapp}
and \citealt{gradshteyn07}, Eq (6.281.1)] to be
\begin{equation}
   V = \frac{v_0}{2\kp} (1-2\beta/3).
   \label{eq:transp}
\end{equation}
Provided the transport and the surface Stokes drift are known, as is usually
the case with wave models, we can now use the assumption that the Phillips
spectrum is a good representation of the Stokes drift to determine an inverse
depth scale $\overline{k}$ by substituting it for the peak wavenumber $\kp$
in \Eq{transp},
\begin{equation}
   \overline{k} = \frac{v_0}{2V} (1-2\beta/3).
   \label{eq:kerf}
\end{equation}
Note that we still need to estimate $\beta$, which for the Phillips spectrum
is exactly one.

\section{Parametric spectra}
\label{sec:paramspec}
We now test the profile (\ref{eq:vz}) on a range of other parametric spectra.
In each case we have estimated $\beta$ by averaging over the range from the
peak frequency $\omp$ to a cut-off frequency here set at $\omega_\mathrm{c} =
10\omp$.

Table~\ref{tab:stats} summarizes the normalized rms (NRMS) error of the
Phillips profile approximation and the previously studied exponential
integral profile.  The NRMS is defined as the difference between the speed
of the approximate profile (mod) and the speed of the full profile, divided
by the transport (which is numerically integrated from the full profile),
\begin{equation}
   \delta v = V^{-1} \int_{-H}^0 |v_\mathrm{mod}-v| \, \dr z.
\end{equation}
Here $H$ is some depth below which the Stokes drift can be considered
negligibly small.

We first compare the Phillips spectrum against the Phillips
approximation. Here, $\beta = 1$ and any discrepancy in terms of NRMS is
due to roundoff error. We then investigate the fit to the Pierson-Moskowitz
(PM) spectrum \citep{pie64} for fully developed sea states,
\begin{equation}
   F_\mathrm{PM}(\omega) = 
     \alp g^2 \omega^{-5} \exp
     \left[-\frac{5}{4}\left(\frac{\omega_\mathrm{p}}{\omega}\right)^4\right].
     \label{eq:PM}
\end{equation}
As seen in Table~\ref{tab:stats}, the NRMS under the PM spectrum is markedly
reduced with the new profile. The $\beta$ value is also quite close to unity.
This is also the case for the JONSWAP spectrum \citep{has73}, with a peak
enhancement factor $\gamma = 3.3$,
\begin{equation}
   F_\mathrm{JONSWAP}(\omega) = F_\mathrm{PM} \gamma^\Gamma,
     \label{eq:JONSWAP}
\end{equation}
where
\begin{equation}
   \Gamma =
   \exp \left[-\frac{1}{2}\left(\frac{\omega/\omp-1}{\sigma}\right)^2\right].
     \label{eq:Gamma}
\end{equation}
Here $\sigma$ is a measure of the width of the peak.  We have also looked
at bimodal, unidirectional spectra by adding a narrow Gaussian spectrum
representing 1.5 m swell at 0.15 Hz and 0.05 Hz to a JONSWAP and PM wind sea
spectrum, respectively. We see in Table~\ref{tab:stats} that the estimates
of $\beta$ for the combined swell and wind sea spectra are still close to
unity. The NRMS difference is markedly higher for the exponential integral
profile proposed by BJB for all spectra, including the bimodal ones.

\begin{figure}[h]
\begin{center}
 \begin{tabular}{cc}
 \hspace{-1.5cm}\includegraphics[scale=0.4]{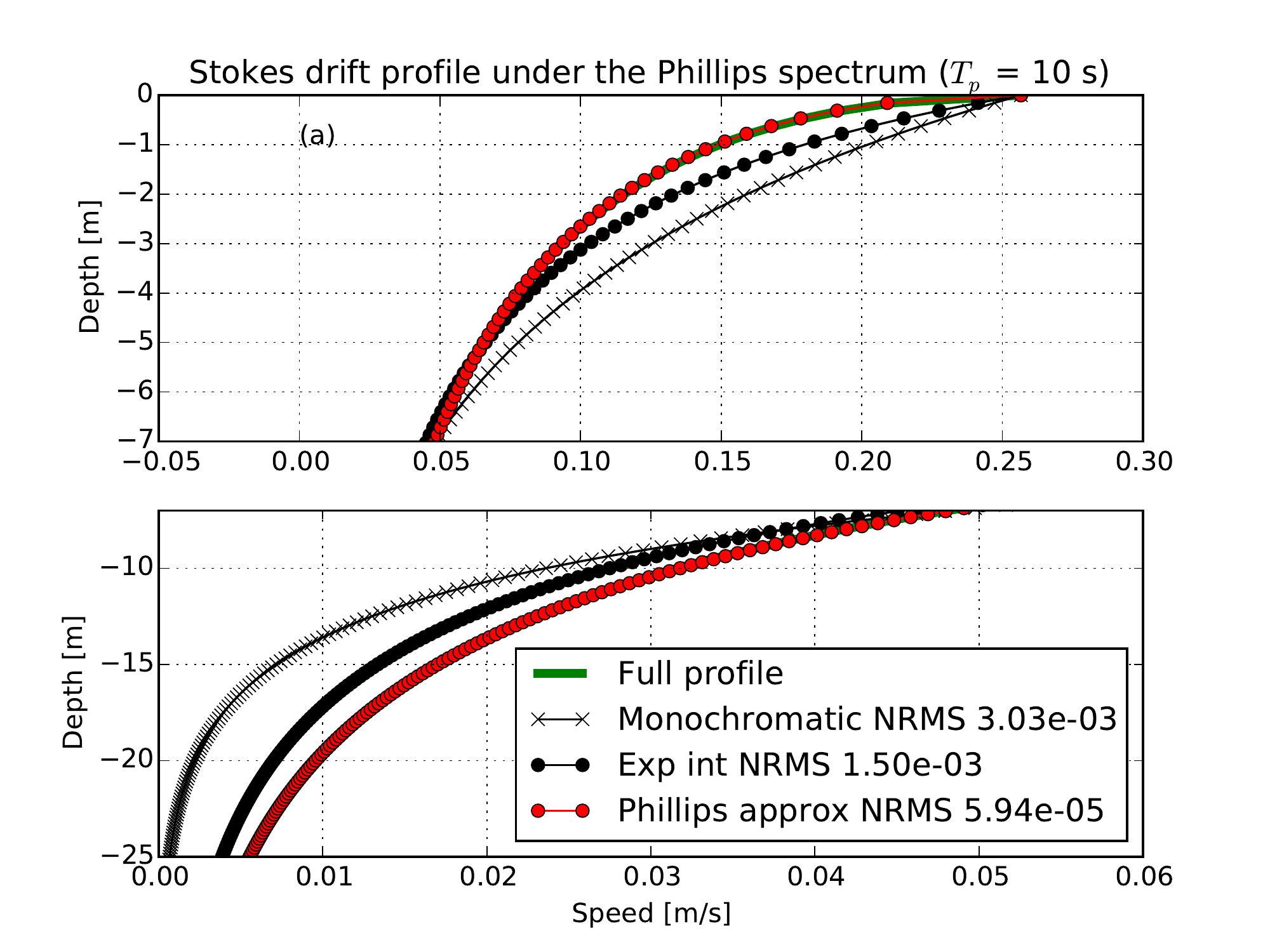} & \includegraphics[scale=0.4]{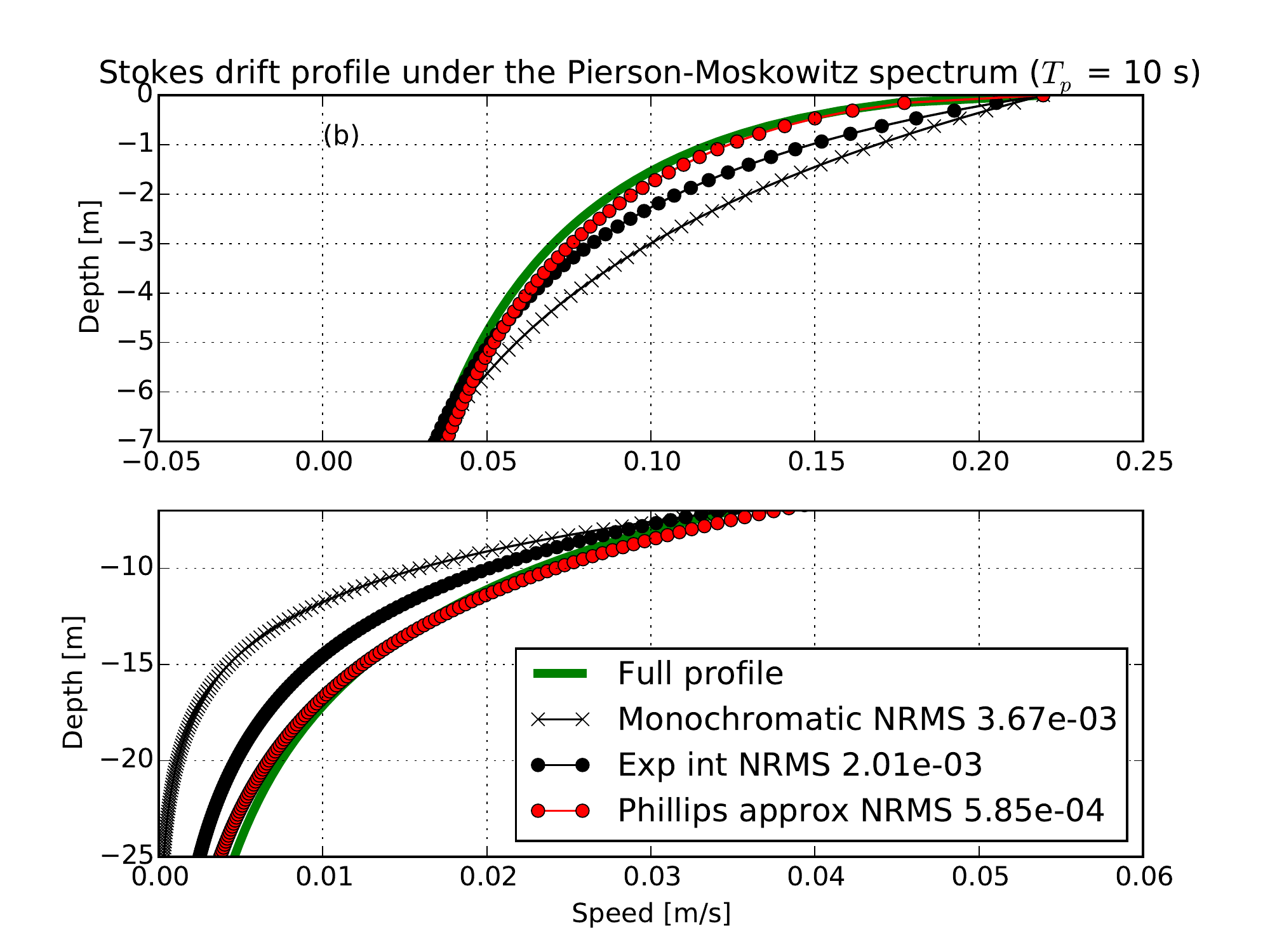}\\
 \hspace{-1.5cm}\includegraphics[scale=0.4]{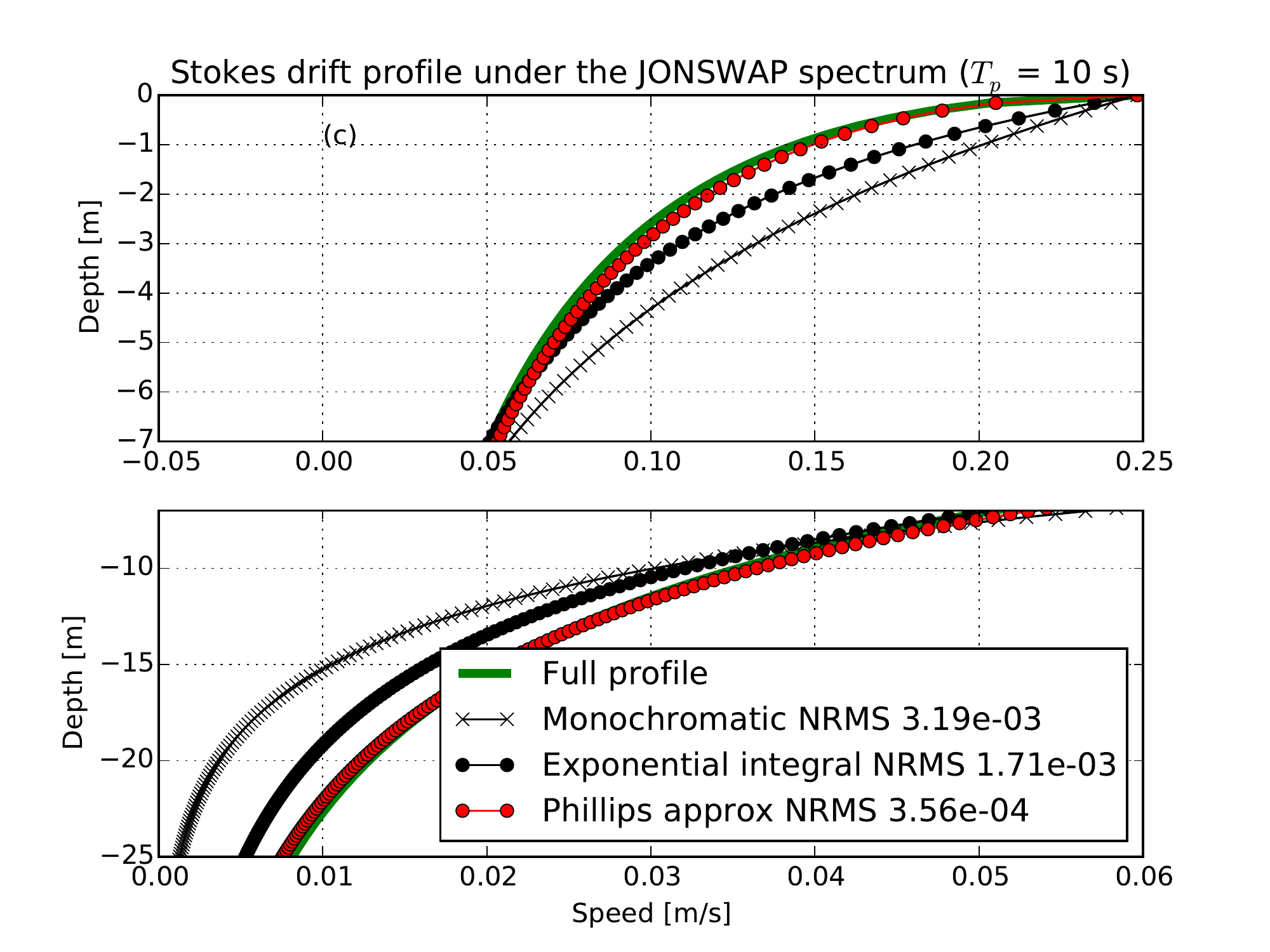} & \includegraphics[scale=0.4]{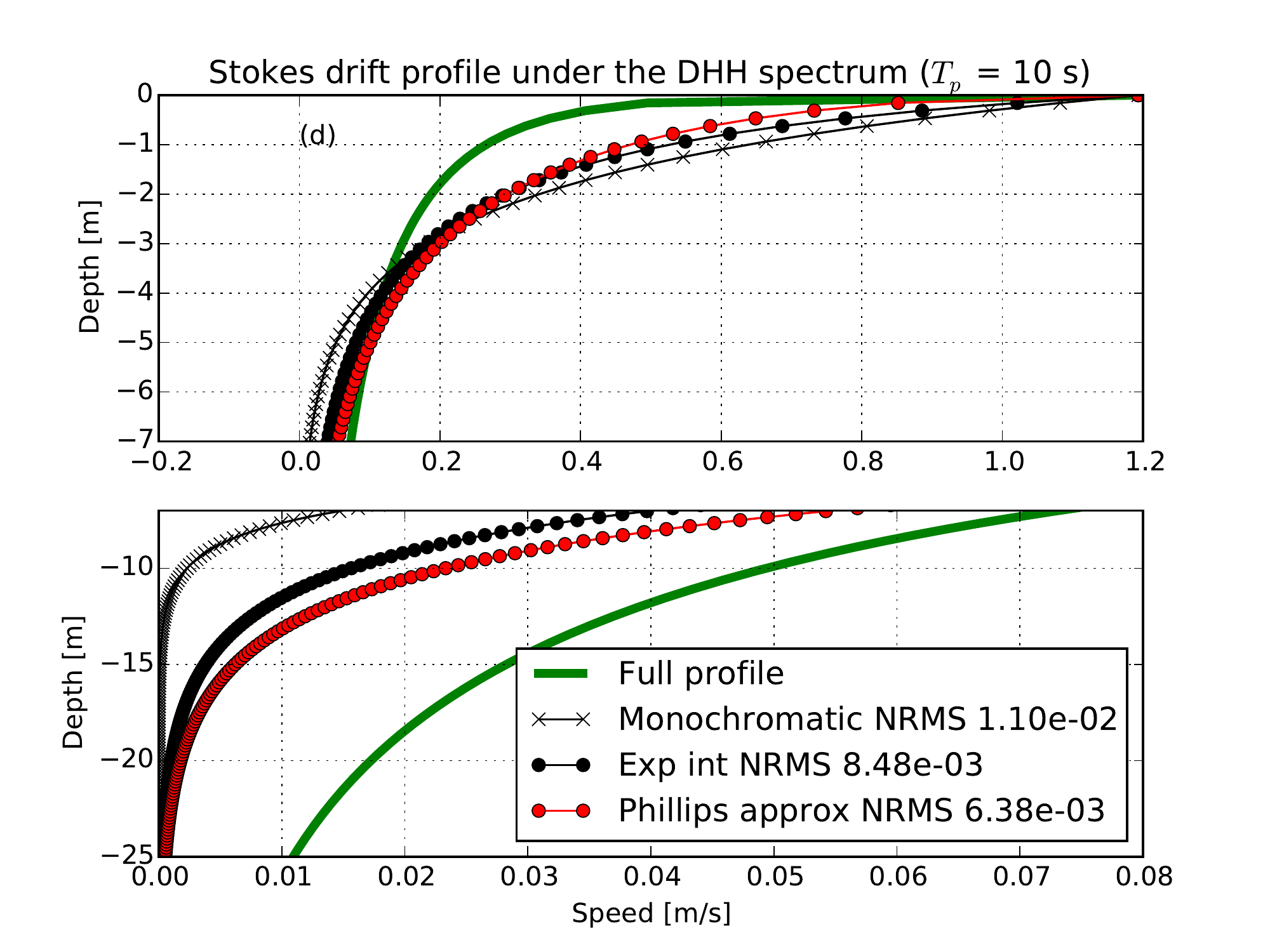}
 \end{tabular}
\caption{A comparison of the merits of the three approximate profiles against four parametric spectra. The normalized rms difference compared to the Stokes profile integrated from the parametric spectrum is marked in the legends. Panel a: The Phillips spectrum. The Phillips approximation is identical to the parametric spectrum to within roundoff error and overlaps exactly (Phillips approximation marked in red; the original Phillips profile in green but underneath the red curve). Panel b: The Pierson-Moskowitz spectrum. Panel c: The JONSWAP spectrum. The Pierson-Moskowitz and JONSWAP spectra are extremely well modelled by the Phillips approximation and overlap nearly perfectly. Panel d: The Donelan-Hamilton-Hui spectrum. This spectrum has an $\omega^{-4}$ and has a quite different Stokes drift profile. The Phillips approximation is still the best of the three approximate profiles.}
\label{fig:paramprof}
\end{center}
\end{figure}

The assumption that the Phillips profile is a good fit to parametric
spectra can also be tested in a more straightforward manner without making
any assumption of the behaviour of the quantity $\omega G(\omega)$ by simply
fitting a Phillips profile ($\beta = 1$) to various spectra. In \Fig{paramprof}
we have fitted the Phillips profile to the surface Stokes drift $v_0$ and
the transport $V$ from  parametric spectra and compared the approximate
profile to the full profile. The results show that for the Phillips spectrum
the approximation matches the full profile (to within roundoff error). More
interestingly, the Pierson-Moskowitz and the JONSWAP spectra are both very
well represented by the Phillips approximation (see \Fig{paramprof}). This
simply confirms what we found in  Table~\ref{tab:stats}. A more challenging
case is the Donelan-Hamilton-Hui (DHH) spectrum \citep{don85} which has an
$\omega^{-4}$ tail,
\begin{equation}
  F_\mathrm{DHH}(\omega) = \alp g^2 \omega^{-4} \omp^{-1}\er^{-(\omp/\omega)^4}\gamma^\Gamma,
  \label{eq:dhh}
\end{equation}
and will consequently behave very differently in the tail.  The spectrum
is identical to the JONSWAP spectrum except for substitution of the peak
frequency $\omp$ for $\omega$ and a Jacobian transformation removing the
factor $5/4$ in the exponential. It is worth noting that the surface Stokes
drift under the DHH spectrum is ill-defined \citep{webb11,webb15}, since
\begin{equation}
  v_\mathrm{DHH}(0) = \alp g^2  \omp^{-1}\int_0^\infty \omega^{-1}\er^{-(\omp/\omega)^4}\gamma^\Gamma \, 
  \dr \omega,
  \label{eq:vsdhh}
\end{equation}
which is unbounded because the integrand asymptotes to
\begin{equation}
 \lim_{\omega \to \infty} \omega^{-1}\er^{-(\omp/\omega)^4}\gamma^\Gamma(\omega) = \omega^{-1}.
 \label{eq:vsdhhinfty}
\end{equation}
Setting a cut-off frequency at $100\omp$ yields the results shown in
\Fig{paramprof} for $T_ \mathrm{p} = 10$ s. As can be seen the Phillips
approximation is not good, but it does in fact represent a small improvement
compared with the monochromatic and exponential integral approximations.

\section{ERA-Interim spectra in open-ocean conditions} \label{sec:era}
\begin{figure}[h]
\begin{center}
  \includegraphics[scale=0.6]{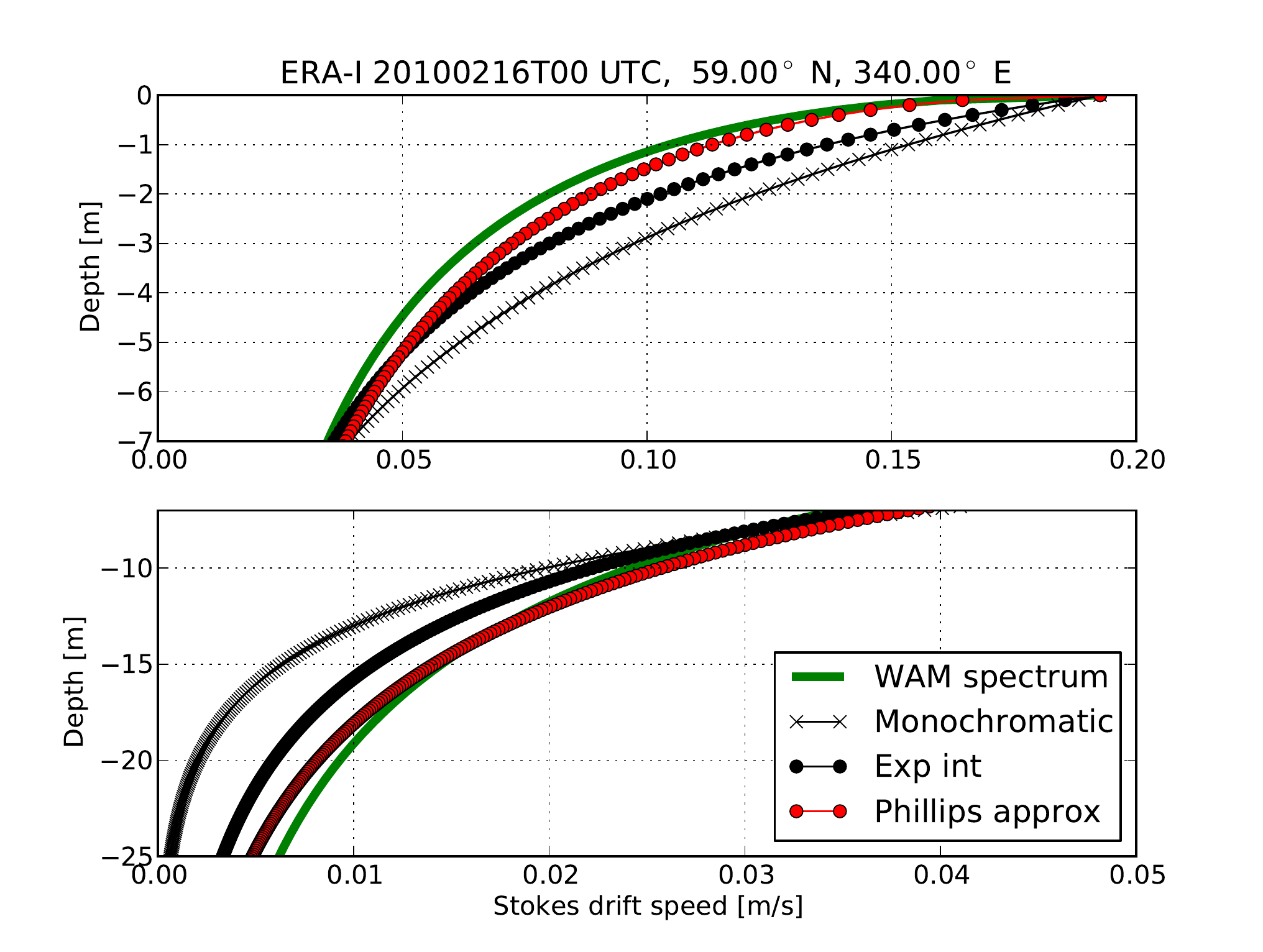}
\caption{The Stokes drift profile under a full two-dimensional wave spectrum
from the ERA-Interim reanalysis. The location is in the north Atlantic. The upper panel is a zoom
of the upper 7 m while the lower panel shows the profile to 25 m. The red line is the Phillips approximation.}
\label{fig:erai_profile}
\end{center}
\end{figure}

\begin{figure}[h]
\begin{center}
 \includegraphics[scale=0.6]{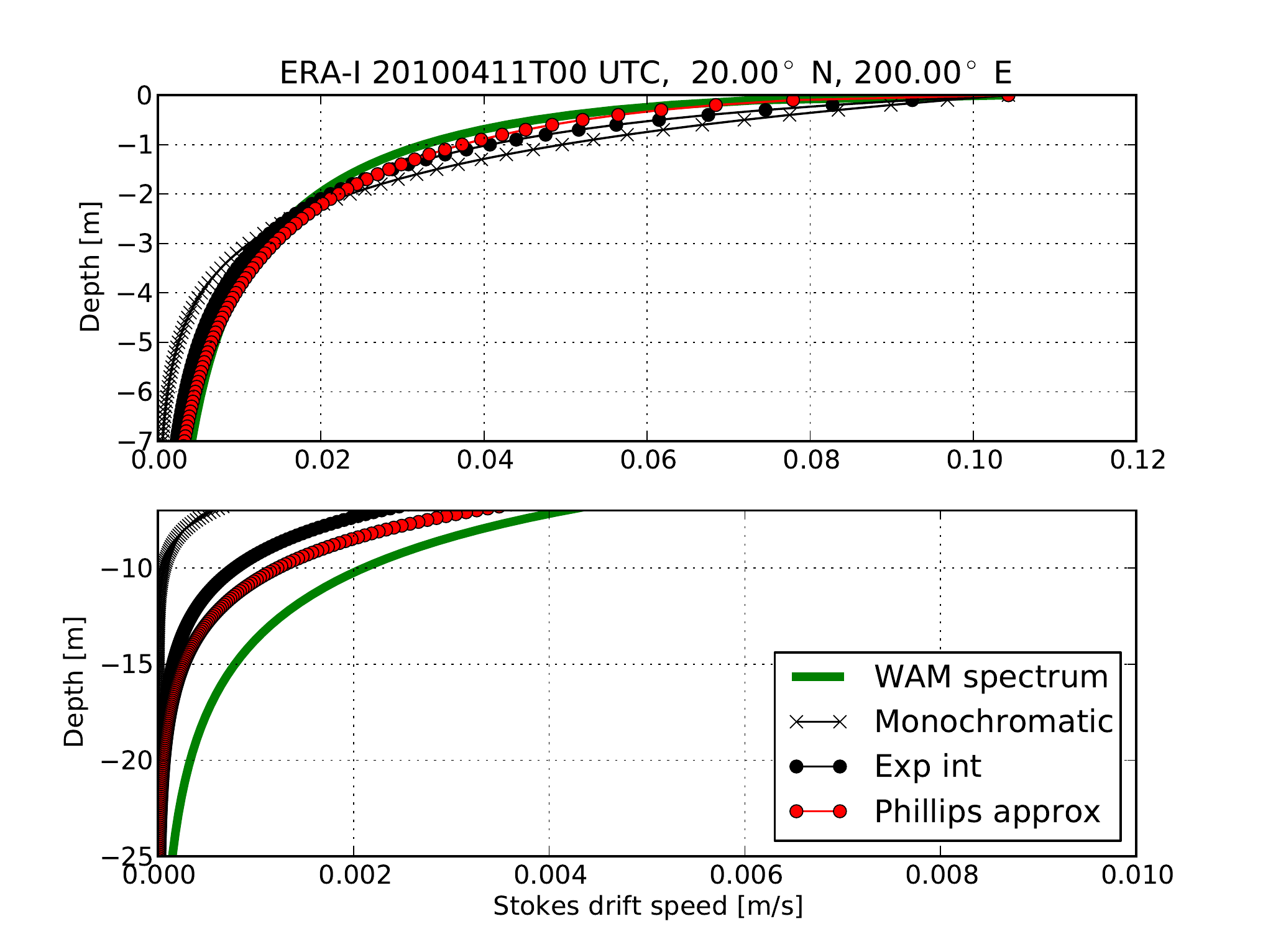}
\caption{The Stokes drift profile under a full two-dimensional wave spectrum
from the ERA-Interim reanalysis. The location is near Hawaii. The upper panel is a zoom
of the upper 7 m while the lower panel shows the profile to 25 m. The red line is the Phillips approximation.}
\label{fig:erai_profile_hawaii}
\end{center}
\end{figure}

\begin{figure}[h]
\begin{center}
 \includegraphics[scale=0.6]{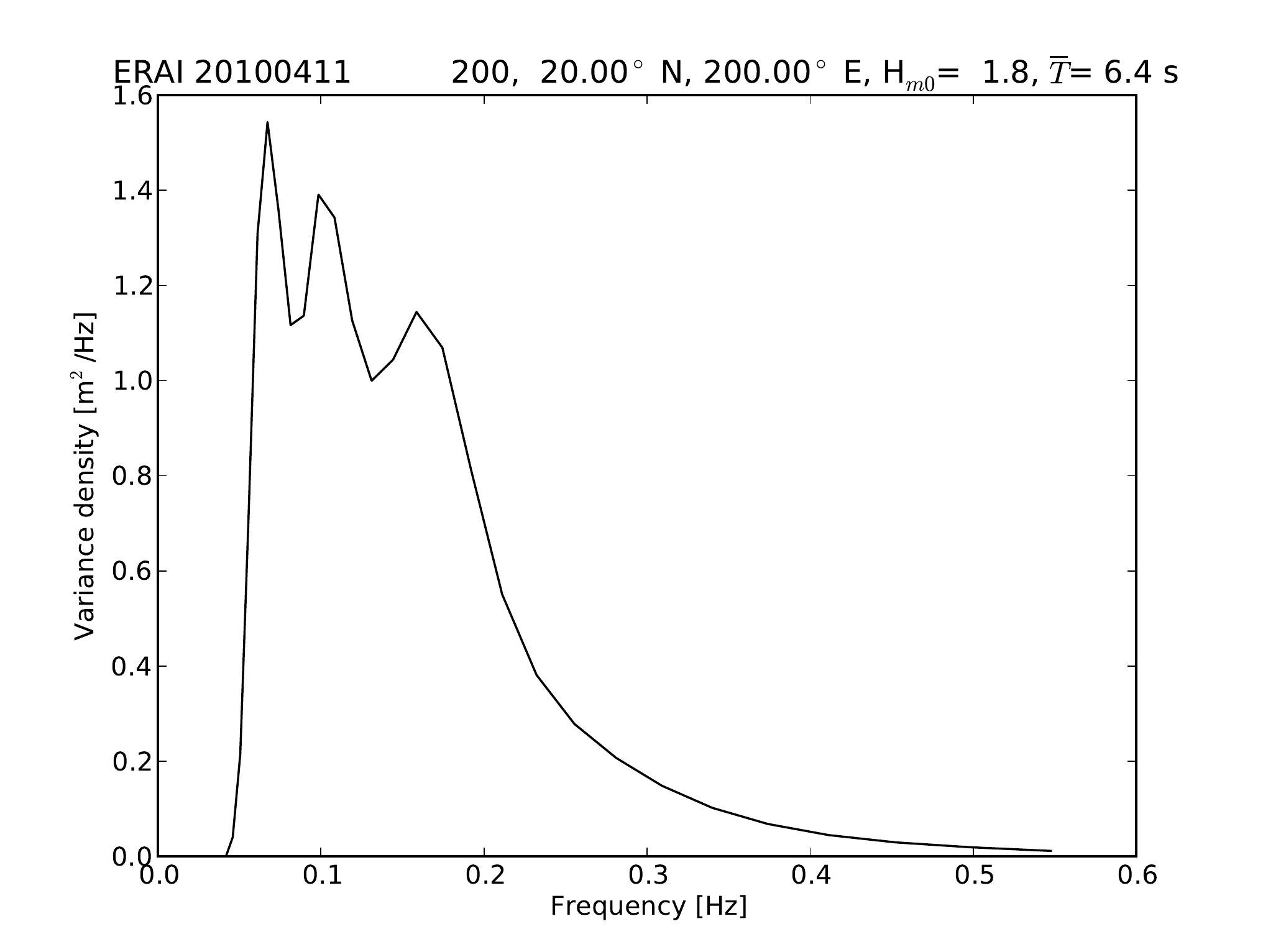}
\caption{
The one-dimensional spectrum associated with \Fig{erai_profile_hawaii}b
shows three peaks corresponding to swell and wind sea.}
\label{fig:erai_spec_hawaii}
\end{center}
\end{figure}

\begin{figure}[h]
\begin{center}
 \includegraphics[scale=0.6]{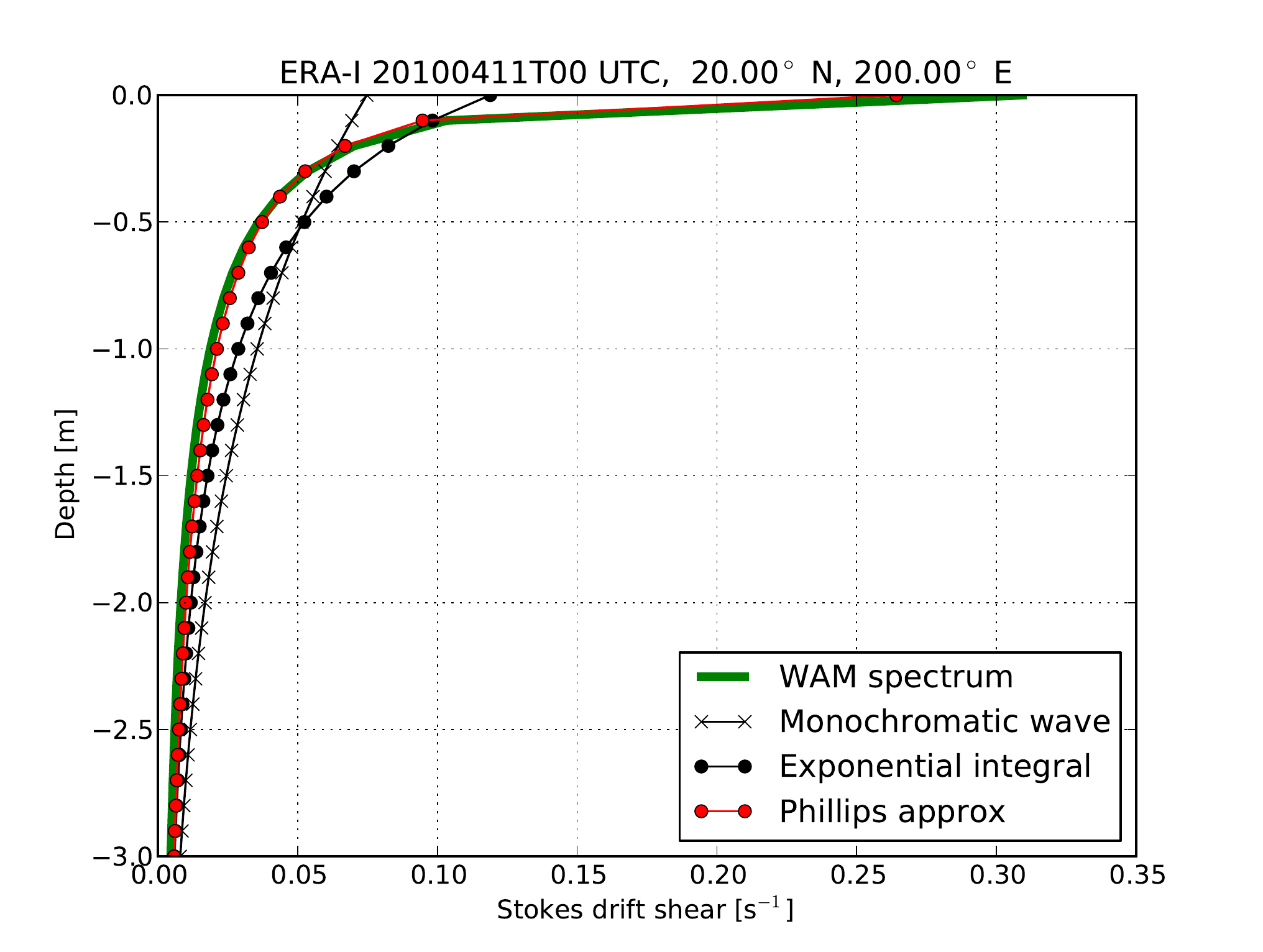}
\caption{The Stokes drift shear under a full two-dimensional wave spectrum
from the ERA-Interim reanalysis. The location is in the swell-dominated
Pacific near Hawaii at 20$^\circ$N, 200$^\circ$E. The red line is the Phillips approximation.}
\label{fig:erai_shear}
\end{center}
\end{figure}

Although $\beta$ can be estimated from the spectrum as shown in \Eq{hatbeta},
it is a quantity which will not be generally available from wave models.
We find that $\overline{\beta} = 1.0$ is a very good approximation for a
dataset of two-dimensional spectra taken from the ERA-Interim reanalysis
\citep{dee11} in the North Atlantic Ocean for the period of 2010 (same location
as used by BJB) as well as a swell-dominated location near Hawaii ($20^\circ$N,
$160^\circ$W). The temporal resolution is six hours and the spatial resolution
of the wave model component of ERA-Interim is approximately 110 km. The angular
resolution is $15^\circ$ while the frequency resolution is logarithmic over 30
frequency bins from $0.0345$ Hz. We have computed the  two-dimensional Stokes
drift velocity vector at every 10 cm from the surface down to 30 m depth from
the full spectra. Comparing the approximate profiles to the full profiles
(see Figs \ref{fig:erai_profile}-\ref{fig:erai_profile_hawaii}) reveals that
in most cases the Phillips profile (\ref{eq:vz}) is a closer match to the full
profile than the exponential integral profile (\ref{eq:ve}), even in cases with
very complex spectra (see the tri-peaked spectrum in \Fig{erai_spec_hawaii}
associated with the profile in \Fig{erai_profile_hawaii}b). In particular,
it is a very good match to the shear near the surface, which becomes very
high, and in the case of the Phillips spectrum infinite.  \Fig{erai_shear}
reveals the much stronger shear near the surface achieved by the Phillips
profile. In fact, the gradient is an almost perfect match to that of the
full profile. This is unsurprising since near the surface the high-frequency
$\omega^{-5}$ tail will dominate the shear.  ECWAM adds a high-frequency
diagnostic tail \citep{wam40r1} \begin{equation}
   \mathbf{v}_\mathrm{HF}(z) = \frac{16\pi^3}{g}f_\mathrm{c}^5
   \int_0^{2\pi} F(f_\mathrm{c},\theta) \hat{\mathbf{k}} \, \dr\theta
   \int_{f_\mathrm{c}}^{\infty} \frac{\exp{(-\mu f^2)}}{f^2} \,\dr f,
   \label{eq:utail}
\end{equation} where $\mu = -8\pi^2z/g$.  This integral is similar to
(\ref{eq:vphil}) and the solution is similar to (\ref{eq:vz})  [see eg
\citealt{gradshteyn07}, Eq~(3.461.5)], yielding \begin{equation}
   \mathbf{v}_\mathrm{HF}(z) = \frac{16\pi^3}{g}f_\mathrm{c}^5
    \int_0^{2\pi} F(f_\mathrm{c},\theta) \hat{\mathbf{k}} \, \dr\theta
    \left[\frac{\exp{(-\mu f_\mathrm{c}^2)}}{f_\mathrm{c}} -
     \sqrt{\mu\pi}\,\erfc\left(f_\mathrm{c}\sqrt{\mu}\right)\right].
     \label{eq:uhf}
\end{equation} For the surface Stokes drift this simplifies to \begin{equation}
   \mathbf{v}_\mathrm{HF}(0) = \frac{16\pi^3}{g}f_\mathrm{c}^4
    \int_0^{2\pi} F(f_\mathrm{c},\theta) \hat{\mathbf{k}} \, \dr\theta.
     \label{eq:uhf0}
\end{equation} Here, the cut-off frequency $f_\mathrm{c}$ of ECWAM is related
to the mean wind sea frequency as $2.5\overline{f}_\mathrm{ws}$.  \Eq{utail}
is exactly the profile under the Phillips spectrum (\ref{eq:phil}) on which
our approximation is based and it is unsurprising then that the profile
(\ref{eq:vphilsolved}) is a good match to the full profile as we get close
to the surface where the high frequency part of the spectrum dominates the
Stokes drift velocity.

\begin{figure}[h]
\begin{center}
\includegraphics[scale=0.7]{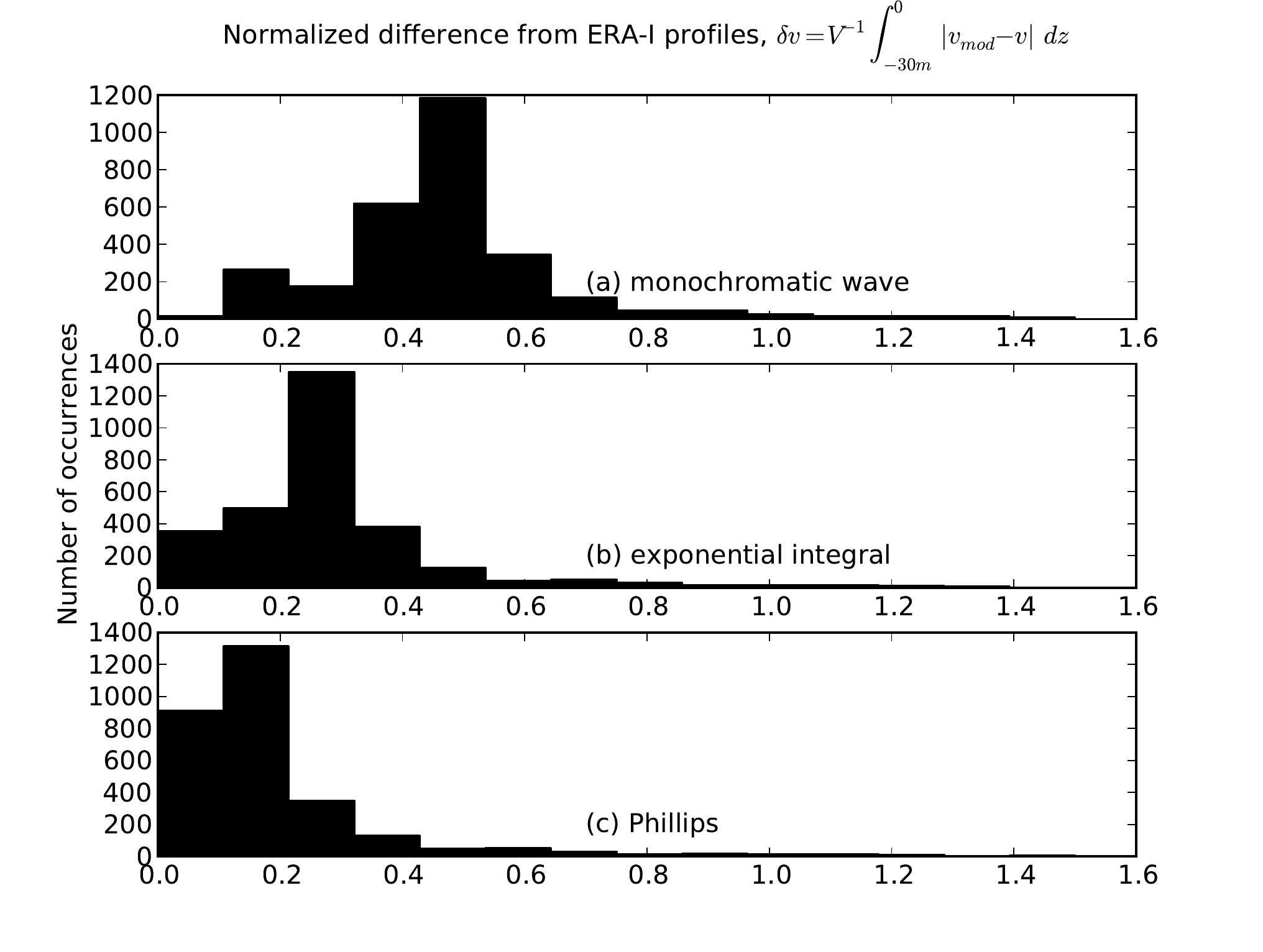}\\
\caption{The NRMS difference between the full Stokes
profile and the monochromatic profile to 30 m depth (vertical resolution
0.1 m). The location is in the North Atlantic.
Panel b: The NRMS difference of the exponential integral profile is
on average about one third that of the monochromatic profile shown in Panel a.
Panel c: The NRMS difference between the Phillips approximation and the full 
profile is about half that of the exponential integral profile (BJB).}
\label{fig:stats_hist}
\end{center}
\end{figure}

\Fig{stats_hist} shows that the Phillips profile has an NRMS deviation
about half that of the exponential integral profile for the North Atlantic
location. The numbers are quite similar for the Hawaii swell location.

\section{Discussion and concluding remarks}
\label{sec:conc}
Although the exponential integral profile proposed by BJB represents a major
improvement over the monochromatic profile, it appears clear that the Phillips
profile (\ref{eq:vphil}) is a much better match, especially for representing
the shear near the surface; see \Eq{philshear2}.  Studies of ERA-Interim
spectra at two open-ocean locations near Hawaii and in the North Atlantic
Ocean show that $\overline{\beta} = 1.0$ is a very good estimate for a wide
range of sea states. This allows us to compute the profile from the same two
parameters as the monochromatic profile, namely the transport and the surface
Stokes drift velocity, and it is thus no more expensive to employ in ocean
modelling. We have shown here that the profile works remarkably well in a
variety of situations, including swell-dominated cases. In \ref{sec:obsapp}
it is shown that the profile is also a better match for profiles under
measured 2~Hz spectra in the central North Sea. This shows that  the fit is
not dependent on the assumption of an $\omega^{-5}$ tail since these spectra
have no high-frequency diagnostic tail added to them. The new profile also
comes closer to the DHH spectrum which has an $\omega^{-4}$ tail, but here
the match is naturally quite poor (see \Fig{paramprof}). We conclude that for
applications concerned with the shear of the profile, in particular studies
of Langmuir turbulence, the proposed profile is a much better choice than
the monochromatic profile, but it is also clearly a better option than the
previously proposed exponential integral profile.

The question of how best to represent a full two-dimensional Stokes drift
velocity profile with a one-dimensional profile was discussed by BJB where
it was argued that using the mean wave direction is better than using the
surface Stokes drift direction since the latter would be heavily weighted
toward the direction of high-frequency waves. This still holds true, but it
is clear that spreading due to multi-directional waves affects the Stokes drift [see \citealt{webb15}],
and although we model the average profile well, situations with for example
opposing swell and wind waves will greatly modify individual profiles. This
will also affect the Langmuir turbulence as parameterised from the Stokes
drift velocity profile, as demonstrated by \citet{mcwilliams14} for an
idealised case of swell and wind waves propagating in different directions.
\citet{li15} investigated the impact of wind-wave misalignment and Stokes
drift penetration depth on upper ocean mixing Southern Ocean warm bias with
a coupled wave-atmosphere-ocean earth system model and found that Langmuir
turbulence, parameterized using a $K$-profile parameterization \citep{large94}.
They found a substantial reduction in the demonstrated that a K-profile
parameterization  for a coupled system consisting of a spectral wave model
and the Community Earth System Model.  This is impossible to model with a
simple parametric profile like the one proposed here, but a combination of
two such parametric profiles, one for the swell and one representing the
wind waves is straightforward to implement.

The method presented here to derive an approximate Stokes drift profile
based on the Phillips profile could also be relevant for other wave-related
processes.  The proposed mixing by non-breaking waves \citep{babanin06} was
implemented in a climate model of intermediate complexity by \citet{babanin09b}
and was compared against tank measurements by \citet{babanin09}.  In a
similar vein, mixing induced by the wave orbital motion as suggested
by \citet{qiao04} has been tested for ocean general circulation models
\citep{qiao10,huang11,fan14}. These suggested mixing parameterizations bear
some semblance to the Langmuir turbulence parameterization in that they involve
the shear of an integral of the wave spectrum with an exponential decay term.
\citet{qiao04} proposes to enhance the diffusion coefficient by adding a
term which involves the second moment of the wave spectrum. It will thus
be somewhat less sensitive to the higher frequencies than the Stokes drift
velocity profile. By again assuming that the wave spectrum is represented
by the Phillips spectrum (\ref{eq:phil}), we find an analytical expression
for the mixing coefficient (see \ref{sec:qiaoapp}). Although we do not pursue
this any further here it is worth noting that similar approximations to those
presented for the Stokes drift profile could thus be found for the proposed
wave-induced mixing by \citet{qiao04}.

Wave-induced processes in the ocean surface mixed layer have long been
considered important for modelling the mixing and the currents in the upper
part of the ocean. Using the proposed profile for the Stokes drift velocity
profile is a step towards efficiently parameterising these processes. Although
more work is needed to quantify the impact of these processes on ocean-only and
coupled models, it appears clear that the impact on the sea surface temperature
(SST) may be on the order of 0.5 K \citep{fan14,janssen13,bre15b}. As the
coupled atmosphere-ocean system is sensitive to such biases, for instance
through the triggering of atmospheric deep convection, see \citet{sheldon14},
wave-induced mixing could play an important role in improving the performance
of coupled climate and forecast models.

\section*{Acknowledgment}
This work has been carried out with support from the European Union FP7
project MyWave (grant no 284455). Thanks to the three anonymous reviewers
and editor Will Perrie for detailed and constructive comments that greatly
improved the manuscript.

\begin{table}[h] 
\begin{center} 
\begin{tabular}{|l|l|l|l|} \hline Spectral shape    & $\beta$ & NRMS Phillips & NRMS exp int \\ \hline
Phillips      &  1  &  0.001 & 0.573 \\ 
JONSWAP ($\gamma=3.3$) &  0.96 & 0.148 & 0.650 \\ 
PM &   1.05   &  0.231 & 0.957 \\ 
JONSWAP+swell ($f=0.15$ Hz) &  0.94  & 0.058 & 0.581 \\ 
PM+l.f. swell ($f=0.05$ Hz)  &  1.04 & 0.240  & 0.920  \\ \hline 
\end{tabular} \end{center}
\caption{Statistics of the two Stokes drift velocity profiles for three
parametric unimodal spectra and two bimodal spectra.  In all experiments the
wind sea peak frequency $f_\mathrm{p}=0.1$ Hz. For the two bimodal spectra the
swell wave height is 1.5 m. The swell frequency is listed in the experiment
description (where l.f. stand for low frequency).}
\label{tab:stats} 
\end{table}

\clearpage

\appendix
\renewcommand\thefigure{\thesection.\arabic{figure}}    
\setcounter{figure}{0}

\section{The transport under a Phillips-type spectrum}
\label{sec:philapp}
The Stokes transport under \Eq{vz} is
\begin{equation}
   V = v_0 \int_{-\infty}^0 \!
       \left[
         \er^{2\kp z} - 
         \beta \underbrace{\sqrt{-2\kp \pi z} \, \erfc \left(\sqrt{-2\kp
         z}\right)}_{\mathrm{GR 6.281.1}}
       \right] \, \dr z.
     \label{eq:philtransp}
\end{equation}
The second term 
can be solved 
by applying Eq (6.281.1) of \citealt{gradshteyn07} as follows. Introduce the
variable substitution $x = \sqrt{-z}$ and rewrite the 
second term (marked GR6.281.1) in \Eq{philtransp}
\begin{equation}
   2\sqrt{2\kp \pi} \int_0^\infty x\, \erfc\left(\sqrt{2\kp} x\right)\, \dr x.
     \label{eq:substx}
\end{equation}
We can now introduce $q=3/2$ and $p=\sqrt{2\kp}$ and employ Eq (6.281.1) of
\citealt{gradshteyn07},
\begin{equation}
   \int_0^\infty x^{2q-1} \erfc \, px \,\dr x =
   \frac{\Gamma(q+1/2)}{2\sqrt{\pi}qp^{2q}} =
   \frac{1}{3\sqrt{\pi}(2\kp)^{3/2}}.
     \label{eq:gr62811}
\end{equation}
The full integral (\ref{eq:philtransp}) can now be written
\begin{equation}
   V = \frac{v_0}{2\kp}(1 - 2\beta/3).
     \label{eq:philtranspsolved}
\end{equation}

\section{An analytical expression for the wave-induced mixing coefficient
         of \citet{qiao04}}
\label{sec:qiaoapp}
The wave-induced mixing coefficient proposed by \citet{qiao04} can be written
\begin{equation}
   B_\nu = \overline{l_{3w}^2}
   \Dp{}{z} \left[\underbrace{\int_0^{2\pi}\!\int_0^{\infty}
   \omega^2 \er^{2kz} E(\omega,\theta) \, \dr \omega \, \dr\theta}_I \right]^{1/2},
   \label{eq:bnu}
\end{equation}
where the mixing length $\overline{l_{3w}} $ is assumed proportional to the
wave orbital radius. We assume that the wave spectrum is represented by
the Phillips frequency spectrum (\ref{eq:phil}), which renders
the integral $I$ in \Eq{bnu} as
\begin{equation}
   I = \alp g^2 \int_{\omp}^\infty 
       \omega^{-3} \er^{2\omega^2z/g} \,\dr\omega.
   \label{eq:bnuint}
\end{equation}
After integration by parts and by performing a variable
substitution $u = \omega^2$ a solution to the integral
(\ref{eq:bnuint}) can be found from Eq~(3.352.2) of \citet{gradshteyn07},
\begin{equation}
  I = \frac{1}{2}\alp g^2\left[
        \omp^{-2} \er^{2\omp^2z/g} - 
        \frac{2z}{g}\Ei(2\omp^2z/g)
      \right].
   \label{eq:bnusolved}
\end{equation}

\section{A comparison against measured spectra in the central North Sea}
\label{sec:obsapp}
\begin{figure}[h]
\begin{center}
\includegraphics[scale=0.7]{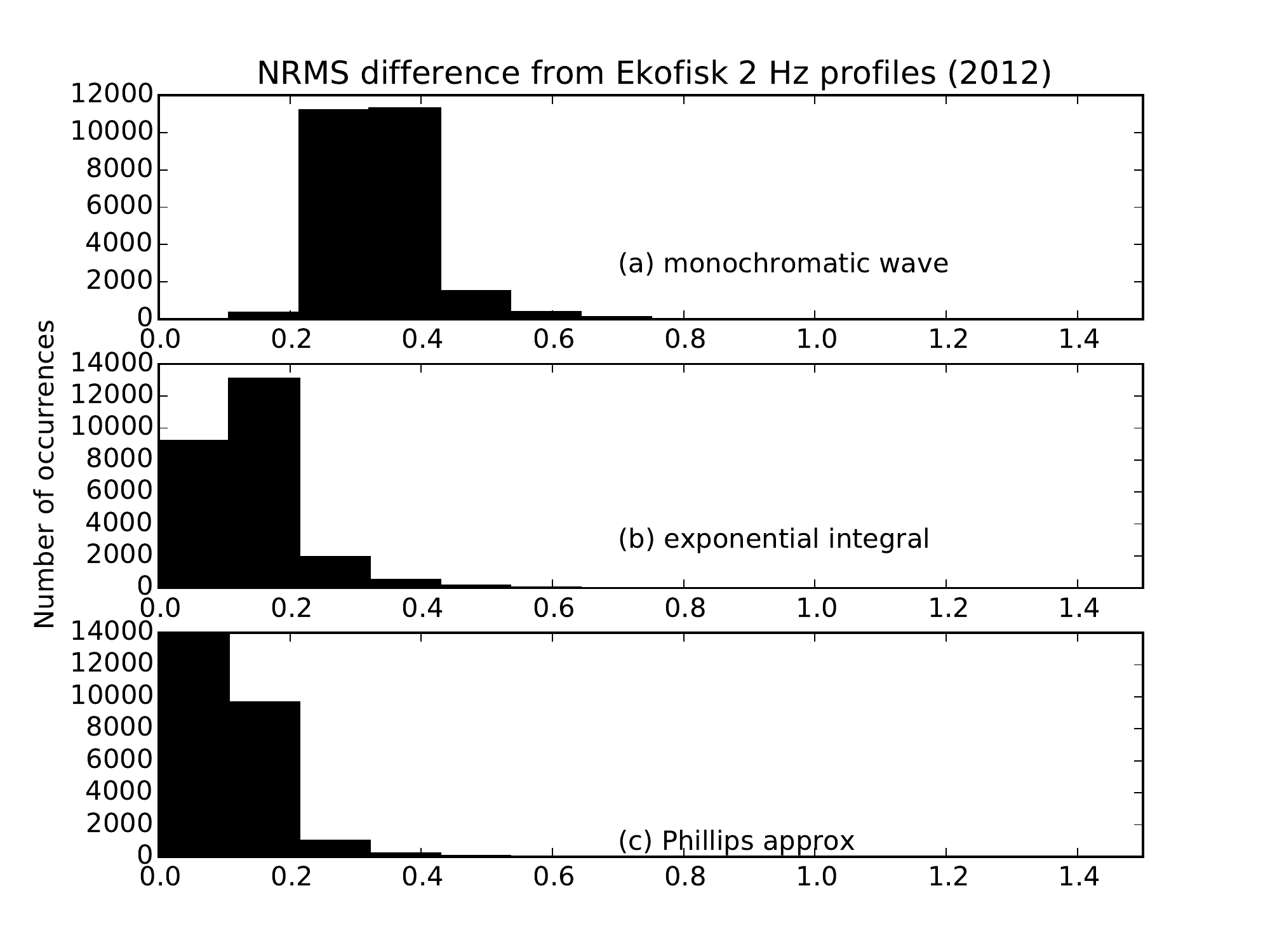}\\
\caption{A comparison of  the full Stokes profile computed from 2 Hz Waverider
observations at Ekofisk ($56.5^\circ$N, $003.2^\circ$E, central North Sea,
72 m depth) for the year 2012 and the three approximate profiles. Panel a:
The average NRMS difference of the monochromatic profile compared to the
full profile is $0.34$.  0.114001530208 Panel b: The NRMS difference of the
exponential integral profile is on average $0.13$ or one third that of the
monochromatic profile shown in Panel a.  Panel c: The NRMS difference between
the Phillips approximation and the full profile is somewhat smaller again
($0.11$).}
\label{fig:stats_hist_obs}
\end{center}
\end{figure}

We have estimated the profile from the same observational spectra as was used
by BJB from the Ekofisk location in the central North Sea for the period
2012 (more than 24,000 spectra in total). The location is ($56.5^\circ$N,
$003.2^\circ$E). The sampling rate was 2 Hz and 20-minute spectra were computed
as described by BJB. The NRMS difference is shown in \Fig{stats_hist_obs}. As
can be seen from Panel c, the new profile reduces the NRMS difference slightly
compared with the exponential integral and quite dramatically compared with
the monochromatic profile. It is worth noting that no $\omega^{-5}$ tail
has been fitted to the spectra, so the improvement is present even without
adding a high-frequency tail.
\clearpage

\end{document}